\documentclass[%
 reprint,
superscriptaddress,
 amsmath,amssymb,
 aps,
prb,
]{revtex4-2}

\usepackage{graphicx}% Include figure files
\usepackage{dcolumn}% Align table columns on decimal point
\usepackage{bm}% bold math
\usepackage[whole]{bxcjkjatype}
\usepackage{color}

\newcommand{\n}{\nonumber \\}
\renewcommand{\t}[1]{\mathrm{#1}}

\def\ket#1{|#1\rangle }
\def\bra#1{\langle #1 |}
\def\braket#1{\langle #1 \rangle}
\def\n{\nonumber \\ }

\begin{document}

%\preprint{APS/123-QED}

\title{
Role of Ward-Takahashi identity in an electron-phonon coupled system \\
--- Revisiting phonon shift current}% 

\author{Takahiro Morimoto}
\affiliation{Department of Applied Physics, The University of Tokyo, Hongo, Tokyo, 113-8656, Japan}

\author{Naoto Nagaosa}
\affiliation{RIKEN Center for Emergent Matter Science (CEMS), Wako, Saitama 351-0198, Japan}
\affiliation{Fundamental Quantum Science Program (FQSP), TRIP Headquarters, RIKEN, Wako 351-0198, Japan}

\date{\today}% It is always \today, today,

\begin{abstract}
We study bulk photovoltaic effects in electron-phonon coupled systems. The conservation of current or gauge invariance, manifested as the Ward-Takahashi identity, plays an essential role in the analysis of the Feynman diagrams, and the leading order contribution to the phonon shift current is identified accordingly.
The leading order contribution essentially arises from the electric polarization carried by optically excited phonons, where the shift current is generated due to a change of electric polarization in the steady state under the optical excitation of phonons.
\end{abstract}

\maketitle

%\tableofcontents

\section{\label{sec:intro}Introduction}
Bulk photovoltaic effect (BPVE) is a focus of recent keen interest from both fundamental physics and applications \cite{ma2021topology,orenstein2021topology,dai2022recent,ma2023photocurrent,Morimoto-JPSJ23,Ahn20,Watanabe21}. 
The quantum geometric nature of nonlinear optical responses is now a rapidly growing research field \cite{Morimoto-JPSJ23,Sipe,Young-Rappe,Morimoto-Nagaosa16}. In particular, the shift current, a representative example of the BPVE, arises from the shift of the wave packet associated with an optical transition, which has a geometric origin. The formulation of the shift current includes the interband transition probability and the corresponding shift vector, which is expressed as the difference in the Berry connections between the conduction and valence bands \cite{Baltz,Sipe,Young-Rappe}. Intuitively, this indicates the change in the electric polarization $P$ during the interband transition, producing the photocurrent according to $J = dP/dt$ \cite{Morimoto-Nagaosa16, Resta24}. This mechanism for the BPVE is in stark contrast with conventional photocurrent generation in p-n junctions, where photo-carriers (electrons and holes) are created and accelerated by a potential gradient, contributing to the net photocurrent. 
Namely, in the case of the shift current, the photocurrent is produced without a bias field, such as a potential gradient that is a necessary ingredient for the p-n junction \cite{Hatada20}. This fact manifests that a photo-carrier itself is not essential for generating the shift current, and rather, the electric polarization associated with the optical excitation is essential. This suggests that quasiparticles that have electronic polarizations can generally produce BPVEs through the shift current mechanism. Indeed, it was revealed that the excitation of excitons \cite{Morimoto-exciton16,Chan21}, magnons \cite{Morimoto-magnon19,Morimoto-magnon21,Hattori24}, and phonons \cite{Okamura22,DielectricLoss24,hu2025optical} can contribute to the shift current without real electron-hole creation.
Even the dielectric loss at the lowest frequency is predicted to be associated with the dc current \cite{DielectricLoss24}.
Experimentally, the dc shift current has been observed by the photoexcitation of excitons \cite{Sotome21,Akamatsu21,Nakamura2024}, 
phonons \cite{Okamura22}, and electromagnons in multiferroic insulators \cite{Ogino2024}.   

In the present paper, we revisit theoretically the shift current due to the phonon excitation, taking into account the Ward-Takahashi (WT) identity, which accounts for the charge conservation or gauge invariance, in the analysis of the Feynman diagrams. 
It turned out that our previous analysis of phonon shift current \cite{DielectricLoss24} was not complete; the diagrams considered in Ref.~\cite{DielectricLoss24}  actually give a subleading contribution in terms of relaxation rate because of partial cancellation as a consequence of the WT identity.  
Instead, we find that the leading order contribution is given by a diagram that essentially measures the electric polarization of optically excited phonons. This result is natural in light of the general mechanism for shift current described by $J = dP/dt$.
Furthermore, the constraint posed by the WT identity on off-resonant electron loop diagrams is directly linked to the nature of the shift current as a polarization response, providing a unified view of the shift current of phonon excitations and enabling its effective theory description.  

The plan of this paper follows. In Sec. \ref{sec:setup}, the setup of the model is introduced. Sec. \ref{sec: phonon SC} is the main body of the paper analyzing the shift current by phonon excitation.
In Sec. \ref{sec: WT id}, the results in Sec. \ref{sec: phonon SC} are discussed from the viewpoint of the Ward-Takahashi identity. The application to a concrete model is given in Sec. \ref{sec: app}, and Sec. \ref{sec: discussion} gives the discussion and conclusions.

\section{\label{sec:setup}Setup}
We describe our setup for electrons in solids coupled to phonon degrees of freedom and an external electric field.

We consider the Bloch electrons in solids described by the Hamiltonian,
\begin{align}
H &= H_\t{el} + H_\t{el-ph}, 
\label{eq: H orbital}
\end{align}
with
\begin{align}
H_\t{el}&=\sum_{k}c_{k}^\dagger h\left(k+\frac{eA}{\hbar}\right) c_{k}, \\
H_\t{el-ph}&= \frac{1}{\sqrt{V}}\sum_{k,q}c_{k+q}^\dagger g\left(k+q+\frac{eA}{\hbar},k+\frac{eA}{\hbar}\right) c_{k} \n
&\qquad \qquad \times (a_q+a_{-q}^\dagger).
\end{align}
Here we consider a multiband electronic system and $h(k)$ is a Bloch Hamiltonian matrix for momentum $k$.
$c_{k}$ represents a vector of annihilation operators for electrons with site/orbital degrees of freedom in the unit cell after the Fourier transform with momentum $k$.
The electromagnetic field $A$ is introduced by the minimal coupling $k \to k+eA/\hbar$ with the electron charge $-e$ ($e>0$).
%For simplicity, we use the convention $e=1$, $\hbar=1$ hereafter.
For a monochromatic electric field $E$ with frequency $\omega$, the electromagnetic field $A$ is given by $A=E/(i\omega)$.
We also introduce electron-phonon coupling $g(k+q,k)$ with an annihilation operator of the phonon mode $a_q$ with momentum $q$.
$g(k+q,k)$ is also a matrix with orbital degrees of freedom and describes the process where an electron with the momentum $k$ is scattered into an electron with the momentum $k+q$ by absorbing the phonon with $q$ (or emitting the phonon with $-q$). 
The $k$ dependence of $g$ comes from the phonon-modulated hopping.
The phonon-modulated hopping process also experiences the presence of the electromagnetic field, which is incorporated with the minimal coupling $k \to k+eA/\hbar$. 
In the following, we focus on spatially uniform excitation of phonons and consider only the $q=0$ mode $a_0 (=a_{q=0})$. We also write  $g(k)=g(k,k)$ for simplicity. 

Let us represent the above Hamiltonian in the band basis.
We consider the energy eigenstates satisfying
\begin{align}
    h(k)u_a(k)=\epsilon_a(k)u_a(k),
\end{align}
with the energy dispersion $\epsilon_a(k)$ and the Bloch wave function $u_a(k)$ for the band $a$ and the momentum $k$.
We write the annihilation operator for the energy eigenstate of the band $a$ as $c_{k,a}$.
The Bloch Hamiltonian part $H_\t{el}$ in Eq.~\eqref{eq: H orbital} is written in the band basis as
\begin{align}
    H_\t{el} &= \sum_{k,a}\epsilon_a(k) c_{k,a}^\dagger c_{k,a} \n
    &+\frac{e}{\hbar}\sum_{k,a,b}
A^\alpha v_{ab}^\alpha(k) c_{k,a}^\dagger c_{k,b} \n
&+ \frac{1}{2}\left(\frac{e}{\hbar}\right)^2 \sum_{k,a,b} A^\alpha A^\beta (\partial_{k_\alpha} v^\beta(k))_{ab} c_{k,a}^\dagger c_{k,b} +O(A^3), 
\end{align}
by expanding the Bloch Hamiltonian $h(k+eA)$ in $A$ and keeping contributions up to $O(A^2)$,
where summation over repeated greek indices is implicitly assumed.
Here $v^\alpha$ is the velocity operator (paramagnetic current operator) along the $\alpha$ direction and is given by $v^\alpha=\partial_{k_\alpha} h$ in the orbital basis.
Similarly, $\partial_{k_\alpha} v^\beta=\partial_{k_\alpha} \partial_{k_\beta} h$ is the diamagnetic current operator.
The operators with subscripts represent their matrix elements as $O_{ab}=\bra{u_a}O\ket{u_b}$.
Generally, the matrix element of a derivative of an operator is obtained by the covariant derivative as
\begin{align}
(\mathcal{D}^\alpha O)_{ab}= \partial_{k_\alpha} O_{ab} - i [\mathcal{A}^\alpha,O]_{ab},
\end{align}
with the Berry connection $\mathcal{A}^\alpha=i\braket{u_a|\partial_{k_\alpha} u_b}$.
In the orbital basis, the covariant derivative is given by the usual $k$ derivative $\mathcal{D}^\alpha O=\partial_k O$.
In particular, the interband matrix element of the Berry connection is written with a matrix element of the velocity operator as
$\mathcal{A}^\alpha_{ab}=-iv_{ab}^\alpha/\epsilon_{ab}$ with $\epsilon_{ab}=\epsilon_a - \epsilon_b$,
which satisfies the above relationship with $v^\alpha=D^\alpha h$. 

Similarly, the electron-phonon part $H_\t{el-ph}$ is expressed in the band basis as
\cite{Mahan}
%[ref: Mahan Eq. (3.200)]
\begin{align}
H_\t{el-ph} =& \frac{1}{\sqrt{V}}\sum_{k,a,b}
\Big[
g_{ab}(k)+
\frac{e}{\hbar}
A^\alpha (\partial_{k_\alpha}g(k))_{ab} \n
&+\frac{1}{2}\left(\frac{e}{\hbar}\right)^2
A^\alpha A^\beta (\partial_{k_\alpha}\partial_{k_\beta}g(k))_{ab} \Big] \n
&\times c_{k,a}^\dagger c_{k,b} (a_0 + a_0^\dagger )
+O(A^3), 
\end{align}
with $g_{ab}(k)=\bra{u_a} g(k) \ket{u_b}$.

The current operator is given by the $A$ derivative of the Hamiltonian as
\begin{align}
    J^\alpha&=-\frac{\delta H}{\delta A^\alpha} \n
    &=-\frac{e}{\hbar}\sum_{k,a,b}
\left[v_{ab}^\alpha(k)+\left(\frac{e}{\hbar}\right) A^\beta (\partial_{k_\alpha} v^\beta(k))_{ab} \right] c_{k,a}^\dagger c_{k,b} \n
&-\left(\frac{e}{\hbar}\right) \frac{1}{\sqrt V}\sum_{k,a,b}
\left[(\partial_{k_\alpha}g(k))_{ab}+\left(\frac{e}{\hbar}\right) A^\beta (\partial_{k_\alpha} \partial_{k_\beta}g(k))_{ab} \right] \n
&\times c_{k,a}^\dagger c_{k,b}(a_0 + a_0^\dagger )
+O(A^2),
\end{align}
We note that the electron-phonon coupling also contributes to the electric current since the phonon-modulated hopping has a coupling to the external electromagnetic field.

In the following, we adopt the convention $e=1$, $\hbar=1$ for simplicity.

\section{Phonon shift current \label{sec: phonon SC}}
In this section, we study the phonon shift current, which is the shift current response arising from the phonon excitation.
Let us consider the current response $J$ of the second-order in the external electric field $E$ as
\begin{align}
    J(\omega_1+\omega_2)&=\sigma(\omega_1+\omega_2;\omega_1,\omega_2)E(\omega_1)E(\omega_2),
\end{align}
where $J(\omega)$ and $E(\omega)$ represent Fourier components of the current density and the electric field, and $\sigma(\omega_1+\omega_2;\omega_1,\omega_2)$ is the second-order conductivity tensor.
The shift current is one mechanism for a DC generation in the second-order electromagnetic response, and is characterized by 
$\sigma(2i\gamma;\omega+i\gamma,-\omega+i\gamma)$.
Here, $\gamma$ is the relaxation rate which is necessary to capture energy dissipation in the photon absorption process.
Since the phonons generally accompany finite electric polarization in the presence of the electron-phonon coupling in inversion broken materials,
optical excitation of phonons leads to an increase in the electric polarization in the system, resulting in finite electric current generation through the shift current mechanism.
We present a diagrammatic formulation of the phonon shift current below.

\begin{figure*}
\begin{center}
\includegraphics[width=0.8\linewidth]{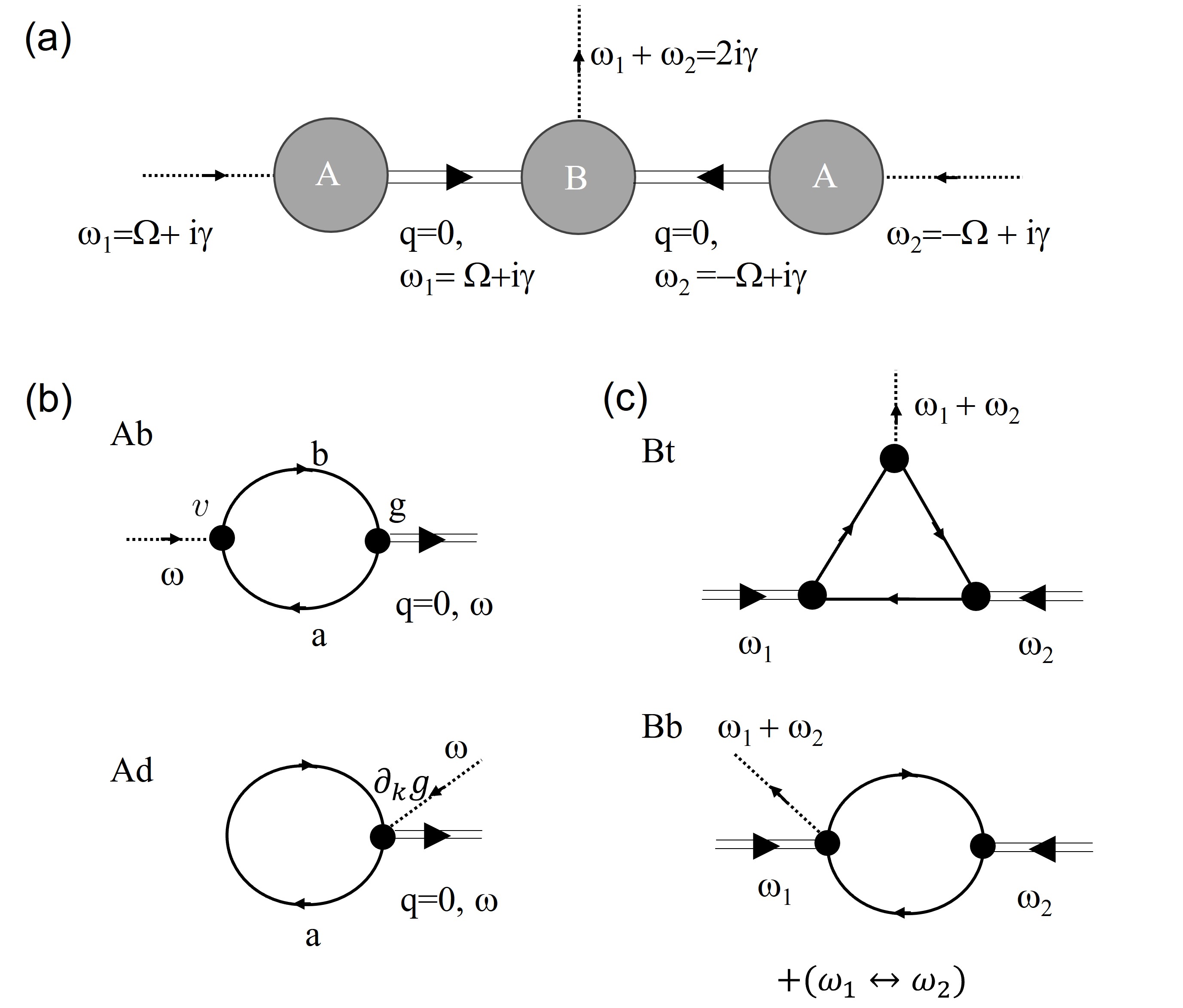}
\caption{\label{fig: diagrams}
Diagrams for phonon shift current.
(a) Diagram for photocurrent response mediated by phonon excitation. 
Left and right electron loops $A$ describe a coupling of a phonon to the external electric field.
The middle electron loop $B$ describes a photocurrent carried by a phonon.
(b) Electron loop diagrams $A$ for the coupling of a phonon to an external electric field.
A bubble diagram $A_b$ and a tadpole diagram $A_d$ contribute to the coupling between the phonon and the light. 
(c) Electron loop diagrams $B$ for the photocurrent carried by a phonon.
A triangle diagram $B_t$ and a bubble diagram $B_b$ contribute to the current response. 
}
\end{center}
\end{figure*}

To describe the phonon shift current, we consider the diagram in Fig. \ref{fig: diagrams}(a). 
Essentially, we consider the optical excitation of phonons as represented by the left and right electron loops $A$ attached to the phonon propagator. 
For the excited phonons, the electric current (i.e., the electric polarization increase) is measured at the middle electron loop $B$.
The second-order current response for the phonon shift current is given by
\begin{align}
    \sigma(\omega_1+\omega_2;\omega_1,\omega_2) &=
    -A(\omega_1)D(\omega_1) B(\omega_1,\omega_2) D(\omega_2)
    A(\omega_2),
    \label{eq: sigma diagram}
\end{align}
which is an expression corresponding to the diagram in Fig. \ref{fig: diagrams}(a).
Here $A(\omega)$ and $B(\omega)$, are diagrams containing an electron loop represented in Fig. \ref{fig: diagrams}(b) and Fig. \ref{fig: diagrams}(c).
$D(\omega)$ is the phonon propagator given by
\begin{align}
    D(\omega)&=\frac{2\omega_0}{\omega^2-\omega_0^2} 
    =
    \frac{1}{\omega-\omega_0} - \frac{1}{\omega+\omega_0}.
\end{align}
$\gamma$ is a small positive number representing the relaxation rate.
We note that the propagation direction of the phonon is interchangeable with $D(\omega)=D(-\omega)$.
The minus sign in Eq.~\eqref{eq: sigma diagram} arises from the current operator in the middle bubble diagram ($B(\omega)$) is given by $-v$ due to our convention of the electron charge, $q=-e$.

%*expressions for A and B in terms of matrix elements. that for sigma

The electron loop diagram $A(\omega)$ describes the phonon creation by the external electric field $E$.
There are two contributions to $A$ as illustrated in Fig.~\ref{fig: diagrams}(b):
a bubble diagram $A_b$ for the paramagnetic current $v$ and the electron phonon coupling $g$,
and modification of the electron-phonon coupling $A_d$ in the presence of the external electric field.
In the imaginary time formulation, the contributions of the diagrams in Fig.~\ref{fig: diagrams}(b) are written as
\begin{align}
    A(i\Omega_1) &= 
    A_b(i\Omega_1) + A_d(i\Omega_1),
\end{align}
with
\begin{align}
    &A_b(i\Omega_1) = \n
    &\frac{1}{i(i\Omega_1)} \int\frac{d\Omega}{2\pi} \int[dk] \t{Tr}
    [v^\beta G(k, i\Omega + i \Omega_1) g G(k, i \Omega)] 
\end{align}
and
\begin{align}
    A_d(i\Omega_1)&=\frac{1}{i(i\Omega_1)} \int\frac{d\Omega}{2\pi} \int[dk] \t{Tr}
    [\partial_{k_\beta} g G(k,i\Omega)].
\end{align}
Here, the prefactors $1/(i(i\Omega_1))$ come from the electromagnetic field $E/(i\omega)$ for the frequency $\omega$.
The electron Green's function $G(k,i\Omega)$ is given in the band representation as
\begin{align}
    G_a(k,i\Omega)&=\frac{1}{i\Omega-\epsilon_a(k)}.
\end{align}
After analytic continuation, the expression for $A(\omega)$ is given by 
\begin{align}
    A(\omega)&=
    \int[dk] \sum_{ab} \frac{g_{ab} v^\beta_{ba}}{i\epsilon_{ba}}I_{2,ab}(\omega),
    \label{eq: A1}
\end{align}
with
\begin{align}
    I_{2,ab}(\omega)&=\frac{f_{ab}}{\omega-\epsilon_{ba}},
\end{align}
where we used the notation $\int [dk]\equiv \int d^dk/(2\pi)^d$.
$f_{ab}=f_a-f_b$ with the Fermi distribution function $f_a$ for the state $a$. The derivation for $A(\omega)$ is detailed in Appendix \ref{app: B(w)}.

The middle electron loop $B$ is expressed as a sum of a triangle diagram contribution $B_t$ and a bubble diagram contribution $B_b$ as
\begin{align}
    B(\omega_1,\omega_2)&=
    B_t(\omega_1,\omega_2)+B_b(\omega_1,\omega_2).
\end{align}
In the imaginary time formulation, the triangle diagram contribution $B_t$ is given by
\begin{align}
    &B_t(i\Omega_1, i\Omega_2) = 
\int \frac{d\Omega}{2\pi} \int[dk] \n
&\times \t{Tr}\Bigg[v^\alpha G(k, i\Omega + i\Omega_1 + i \Omega_2) g G(k, i\Omega + i \Omega_2) g G(k, i \Omega) \n
&+(\Omega_1 \Leftrightarrow \Omega_2)
\Bigg],
\end{align}
and the bubble diagram contribution $B_b$ is given by
\begin{align}
    &B_b(i\Omega_1, i\Omega_2) = 
\int \frac{d\Omega}{2\pi} \int[dk] \n
&\times\t{Tr}\Bigg[\partial_{k_\alpha} g G(k, i\Omega + i \Omega_1) g G(k, i \Omega) 
+(\Omega_1 \Leftrightarrow \Omega_2)
\Bigg],
\label{eq: Bb}
\end{align}
Here the bubble diagram obtained by swapping the two vertices in Fig.~\ref{fig: diagrams} is included in the term $(\omega_1 \Leftrightarrow \omega_2)$ in Eq.~\eqref{eq: Bb}, where the two contributions are Hermitian conjugate to each other.
As detailed in Appendix \ref{app: B(w)}, the corresponding expressions after analytic continuation is given by
\begin{align}
B_t(\omega_1,\omega_2) &=
\int[dk] \sum_{a,b,c\neq a} v^\alpha_{ac} g_{cb} g_{ba} I_{3,abc}(\omega_1, \omega_2)
\n
&+(\omega_1 \Leftrightarrow \omega_2),
\end{align}
with
\begin{align}
I_{3,abc}(\omega_1,\omega_2)
&=\frac{1}{\epsilon_{ac}+\omega_1+\omega_2}\left(\frac{f_{ab}}{\omega_1-\epsilon_{ba}} - \frac{f_{cb}}{-\omega_2 -\epsilon_{bc}}\right),
\end{align}
and
\begin{align}
B_b(\omega_1,\omega_2)
&= \int[dk] \sum_{ab}  (\partial_{k_\alpha} g)_{ab} g_{ba} I_{2,ab}(\omega_1)
+(\omega_1 \Leftrightarrow \omega_2).
\end{align}

As we show in Appendix~\ref{app: B(w)}, $B(\omega_1,\omega_2)$ is proportional to $\gamma$ for $\omega_1=\omega+i\gamma$ and $\omega_2=-\omega+i\gamma$ as
\begin{align}
    B(\omega+i\gamma, -\omega+i\gamma)=\gamma B'(\omega) +O(\gamma^2),
    \label{eq: B(w)}
\end{align}
with
\begin{align}
    B'(\omega) &= 
2i\int[dk] \Bigg\{
\sum_{a,b}(\partial_{k_\alpha}g)_{ab} g_{ba} 
\partial_\omega I_{2,ab}(\omega) \n
&+
\sum_{a,b,c\neq a}v^\alpha_{ac} g_{cb} g_{ba} [\partial_{\omega'}I_{3,abc}(-\omega,\omega')|_{\omega'=\omega} \n
&+ \partial_{\omega'}I_{3,abc}(\omega',-\omega)|_{\omega'=\omega}]
\Bigg\},
\label{eq: B'}
\end{align}
where $B'(\omega)$ is independent of $\gamma$. 
In particular, the diagram $B$ becomes zero for $\gamma=0$, which is a consequence of the Ward-Takahashi (WT) identity as we detail in Sec.~\ref{sec: WT id}.
Namely, the contributions of the diagrams $B$ vanish for $\omega_1+\omega_2=0$ (i.e. $\gamma=0$) and their leading order contribution is proportional to $\gamma$.

Thus, the expression for the phonon shift current is
\begin{align}
    \sigma^\t{ph}(\omega) 
    &\equiv \sigma(2i\gamma;\omega+i\gamma,-\omega+i\gamma) \n
    &= -A(\omega)D(\omega+i\gamma) \gamma B'(\omega) D(-\omega+i\gamma) A(-\omega)
    \label{eq: phonon shift current}
\end{align}
which we call the phonon shift current conductivity in the following.
When the photon frequency is resonant with the phonon frequency $\omega \sim \omega_0$, we can write $\gamma D(\omega+i\gamma) D(-\omega+i\gamma) = \gamma/[(\omega-\omega_0)+\gamma^2] \simeq \pi \delta(\omega-\omega_0)$ (with $\gamma \to 0$), which leads to
\begin{align}
    \sigma^\t{ph}(\omega) &=
    -\pi A(\omega)A(-\omega) B'(\omega) 
    \delta(\omega-\omega_0),
    \label{eq: sigma ph}
\end{align}
showing a resonant enhancement of the phonon shift current at the phonon excitation.
Here, the expression for $B'(\omega)$ is given in Eq.~\eqref{eq: B'}.

\subsection{Expression in low frequency regime ($\omega \ll E_g$)}
When the frequency of incident light $\omega$ is much smaller than the electronic energy scale (typically electronic band gap $E_g$), we can further simplify the expression for the phonon shift current.
The expression for $A(\omega)$ becomes
\begin{align}
    A(\omega=0)&=i\int[dk] \sum_{ab} \frac{f_{ab}g_{ab} v^\beta_{ba}}{\epsilon_{ba}^2},
\end{align}
assuming the interband energy difference $\epsilon_{ba} \gg \omega$.
Similarly, the expression for $B(\omega)$ in the low frequency regime becomes $B(i\gamma, -i\gamma) =\gamma B'(\omega=0)$ with 
\begin{align}
    &B'(\omega=0)= \n
    & 
    2i\int[dk] \Bigg[
    \sum_{a,b}-\frac{f_{ab}(\partial_{k_\alpha}g)_{ab} g_{ba}}{\epsilon_{ba}^2} 
    \n
    &-
    \sum_{a,b,c\neq a}v^\alpha_{ac} g_{cb} g_{ba} 
    \left(\frac{\epsilon_{ac}-2\epsilon_{ba}}{\epsilon_{ac}^2\epsilon_{ba}^2}f_{ab}
    +\frac{\epsilon_{ac}+2\epsilon_{bc}}{\epsilon_{ac}^2\epsilon_{bc}^2}f_{cb}\right) \Bigg].
\end{align}
Also, the terms involving the phonon propagators behave as
\begin{align}
    \tilde D(\omega) &\equiv \gamma D(\omega+ i\gamma) D(-\omega + i\gamma) \n
    &=
    \begin{cases}
        \pi \delta(\omega-\omega_0) & (\omega \simeq \omega_0), \\
        \frac{4\gamma}{\omega_0^2}  & (\omega \simeq 0).
    \end{cases}
    \label{eq: D tilde}
\end{align}
With these expressions, one can represent the phonon shift current conductivity in the low-frequency regime as
\begin{align}
    \sigma^\t{ph}(\omega)& \simeq -A(\omega=0)^2 B'(\omega=0) \tilde D(\omega).
\end{align}
In particular, we obtain
\begin{align}
    \sigma^\t{ph}(\omega)& \simeq -\pi A(0)^2 B'(0) \delta(\omega-\omega_0),
    \label{eq: sigma ph low w}
\end{align}
for the phonon resonance.
In the dc limit, the phonon shift current conductivity of $\sigma^\t{ph}(\omega=0) \propto \gamma$ (from Eq.~\eqref{eq: D tilde}) may seem to imply a nonreciprocal dc transport due to phonons, but the effective theory for phonons presented in Sec.~\ref{sec: effective theory} suggests that such contributions do not give a measurable response.
Namely, while we treated the relaxation rate $\gamma$ as a constant, it is natural that the dissipation does not occur at the equilibrium with $\omega=0$. 
This suggests that the relaxation rate has an $\omega$ dependence as $\gamma=\alpha \omega$ ($\alpha$: constant) around $\omega=0$ and the phonon shift current vanishes as $\sigma^\t{ph}(\omega=0) \propto \gamma \propto \omega$.

In the presence of time reversal symmetry, matrix elements of the velocity operator and the electron-phonon coupling satisfy
\begin{align}
    v_{ab}(k)&=-v_{ab}(-k)^*,\\
    g_{ab}(k)&=g_{ab}(-k)^*,\\
    (\partial_k g)_{ab}(k)&=-(\partial_k g)_{ab}(-k)^*.
\end{align}
Using these properties, one can further reduce the expressions for $A$ and $B$ as
\begin{align}
    A(\omega=0) &=-\t{Im}\left[\int[dk] \sum_{ab} \frac{f_{ab}g_{ab} v^\beta_{ba}}{\epsilon_{ba}^2}\right],
\end{align}
and
\begin{align}
    &B'(\omega=0)= \n
    &
    2\int[dk] \Bigg[
    \sum_{a,b}\frac{f_{ab}\t{Im}[(\partial_{k_\alpha}g)_{ab} g_{ba}]}{\epsilon_{ba}^2} 
    \n
    &+
    \sum_{a,b,c\neq a}\t{Im}[v^\alpha_{ac} g_{cb} g_{ba}] 
    \left(\frac{\epsilon_{ac}-2\epsilon_{ba}}{\epsilon_{ac}^2\epsilon_{ba}^2}f_{ab}
    +\frac{\epsilon_{ac}+2\epsilon_{bc}}{\epsilon_{ac}^2\epsilon_{bc}^2}f_{cb}\right) \Bigg].
\end{align}
From the expressions for $A(\omega=0)$ and $B'(\omega=0)$, the phonon shift current conductivity in Eq.~\eqref{eq: sigma ph low w} becomes proportional to $g_0^4$ with the strength of the electron-phonon coupling $g_0$.
We consider its implications in Sec.~\ref{sec: discussion}.

\subsection{Effective theory for phonons \label{sec: effective theory}}

We consider a low-energy effective theory for phonons.
In the low frequency regime, $\omega \ll E_g$, we can integrate out the electron degrees of freedom and focus on the phonons which are coupled to the external electric field $E$.

The external electric field $E$ enters the original system as a coupling to electric current $J$ with $-(E/i\omega)J$ with an incoming frequency $\omega$.
Thus, phonon's linear coupling to $E$ is given by an electron bubble $A(\omega=0)$ in Fig.~\ref{fig: diagrams}(b).

Phonon's second order coupling to $E$ is obtained from the electron bubble diagram $B(\omega)$ in Fig.~\ref{fig: diagrams}(c). 
A current response $J(\delta \omega)$ is related to the polarization $P(\delta \omega)$ with $P(\delta \omega)=J(\delta \omega)/(-i\delta \omega)$ because of $J=dP/dt$.
The diagram $B(\omega+i\gamma,-\omega+i\gamma)$ describes the current response with outgoing frequency $\delta\omega=2i\gamma$ (which can be reread as an incoming frequency $-\delta \omega=-2i\gamma$).
Thus, the second-order coupling is given by
\begin{align}
    P_2&=\frac{-B(i\gamma,i\gamma)}{-i\delta \omega} =-\frac{1}{2}B'(\omega=0).
\end{align}
Note that the minus sign appears due to the definition of the current operator $J=-ev$.

Thus, the low-energy effective theory for the phonons is given by
\begin{align}
     H_\t{eff}&=\hbar \omega_0 a^\dagger a - E\left[P_1(a_0 + a_0^\dagger) + \frac{P_2}{2} (a_0 + a_0^\dagger)^2 \right],
     \label{eq: Heff for phonon}
\end{align}
with
\begin{align}
    P_1&=-A(\omega=0), & 
    P_2&=-\frac{1}{2}B'(\omega=0).
    \label{eq: P1 P2}
\end{align}
This effective theory reproduces the phonon shift current in the low-frequency regime as follows.
The phonon creation rate $\alpha$ is given by 
\begin{align}
    \alpha(\omega)&=2E^2\pi|P_1|^2\delta(\omega-\omega_0),
\end{align}
according to Fermi's golden rule.
Since each phonon accompanies electric polarization $P_2$,
the electric polarization in the system increases by $\alpha(\omega)P_2$ in time due to phonon creation.
According to $J=dP/dt$, the phonon shift current conductivity is given by
\begin{align}
    \sigma^\t{ph}(\omega)&= 2\pi |P_1|^2 P_2 \delta(\omega-\omega_0) \n
    &= -\pi A(0)^2 B'(0) \delta(\omega-\omega_0),
\end{align}
which reproduces Eq.~\eqref{eq: sigma ph low w} for the low frequency regime.
This clearly indicates that the phonon shift current is driven by the electric polarization of optically excited phonons, which is quite analogous to the case of the shift current of magnons, where the electric polarization of optically excited magnons (electromagnons) drives the photocurrent generation \cite{Morimoto-magnon21}.

In the DC limit $\omega=0$, the effective theory indicates that the equilibrium point for the phonon coordinate is just shifted with nonzero $E$.
Once the new equilibrium with a shifted phonon coordinate is realized, no current response takes place, indicating the photocurrent is zero for $\omega=0$ in an insulating state even in the presence of phonons.

A few words are in order for the role of relaxation.
In the process of shift current generation by phonon excitation, relaxation plays an essential role.
If the phonon excitation decays locally, the electric polarization carried by the phonon also disappears, leading to a back-flow current which cancels the shift current generated by the original phonon excitation.
In the presence of relaxation, such as scattering of phonons or coupling to electrodes, the electric polarization carried by the phonons is not canceled out within the sample, and net current flow is realized.
In particular, to sustain a constant electric current flow, it would be necessary to generate some form of an electronic carrier, where the electric polarization of phonons is transferred to those carriers and retrieved out of the sample through those carriers.
Such carrier generation within the sample can be induced by the pile-up of the electric polarization of the phonon excitation.
The mechanism of carrier generation is generally complicated and may be material-dependent.
In any case, the way the phonons go  through relaxation and dissipation is essential in the actual generation of the shift current transport.
In this regard, the shift current formula, such as Eq.~\eqref{eq: sigma ph}, gives an upper limit of the current which can be retrieved.

\section{Ward-Takahashi identity for photocurrent response \label{sec: WT id}}
In this section, we consider the phonon shift current in terms of the Ward-Takahashi (WT) identity.

In the relativistic case, we consider a correlation function 
$M(k;p_1 \dots p_n; q_1 \dots q_n)=\epsilon_\mu(k)M^\mu(k;p_1 \dots p_n; q_1 \dots q_n)$ 
with $n$ incoming electrons with momenta $(q_1 \dots q_n)$, $n$ outgoing electrons with momenta $(p_1 \dots p_n)$, an incoming photon with energy-momentum $k$, where $\epsilon_\mu(k)$ is the polarization vector ($\mu=0,1 \dots 3$).
The WT identity relates this correlation function to those without the external photon as
\begin{align}
    &k_\mu M^\mu(k;p_1 \dots p_n; q_1 \dots q_n) \n
    & = \sum_i [M_0 (p_1 \dots p_n; q_1 \dots (q_i-k) \dots q_n) \n
    &- M_0 (p_1 \dots (p_i+k) \dots p_n; q_1  \dots q_n)],
\end{align}
where $M_0(p_1 \dots p_n; q_1  \dots q_n)$ is a correlation function with $n$ incoming electrons with momenta $(q_1 \dots q_n)$ and $n$ outgoing electrons with momenta $(p_1 \dots p_n)$ \cite{Ward,Takahashi,Peskin}.

Let us consider a correlation function involving one electron line.
The WT identity states
\begin{align}
    k_\mu M^\mu(k;p; q)
    = \sum_i [M_0 (p; q-k) - M_0 (p+k; q)].
    \label{eq: WT id 1 electron}
\end{align}
In the nonrelativistic case with $k_\mu=(\omega,k_i)$ $(i=1,2,3)$, the photon vertex for $M^0$ is given by just $1$ and the photon vertex for $M^i$ $(i=1,2,3)$ is given by a current operator for the $i$th direction \cite{Mahan,nozieres2018theory}.
Also in a lattice system, $k_i$ in the left-hand side is replaced by terms such as $2\sin k/2$, but Eq.~\eqref{eq: WT id 1 electron} still holds as far as $(\omega, k_i)$ is sufficiently small.   

The WT identity holds even when the correlation function involves incoming/outgoing phonons. Thus, the diagrams $A(\omega)$ and $B(\omega)$ in Fig.~\ref{fig: diagrams}(b,c) obey the WT identity.
%In particular, when the frequency of the external electric field is zero, the left-hand side only involves correlation functions with a current vertex for the electric field.
In addition, when the electron line forms a loop, the two terms on the right-hand side in Eq.~\eqref{eq: WT id 1 electron} become the same expressions just with shifted momenta for the internal electron line ($q \to q-k$ and  $p \to p-k$) which cancel out upon an integration over the momentum $q$.
This leads to the WT identity for loop diagrams,
\begin{align}
    \omega M^0_\t{loop}(k;p;q) - k_i M^i_\t{loop}(k;p;q)=0,
    \label{eq: WTI loop}
\end{align}
where we split the LHS into a frequency term $\propto \omega$ and a momentum term $\propto k_i$.
The vertex operator for $M^0_\t{loop}$ is $1$, while that for $M^i_\t{loop}$ is the current operator $v^i$ along the $i$th direction (including modulation of other input vertices with the incoming photon).
If we choose the momentum $k_\mu$ as $\omega=0$ and $k_i=\delta k \delta_{ij}$ (only the $j$th spatial component is nonzero), the WT identity indicates that $M^j_\t{loop}(k;p;q)=0$.
Thus, a loop diagram $M^j_\t{loop}(0;p;q)$ with a current vertex coupled to an external electric field with zero frequency ($\omega=0$) vanishes as a consequence of the WT identity when one can take the $k_\mu \to 0$ limit.
This also indicates that the loop diagrams coupled to an external electric field with small frequency $\omega$ give $O(\omega)$ contributions, as far as the series expansion of $M^j_\t{loop}(k;p;q)$ in $\omega$ is well-defined. 
We indeed found this is true in the calculation for $A(\omega)$ with small  $\omega$ and $B(\omega)$ with small $\omega_1+\omega_2$ in Appendix \ref{app: B(w)}.

One caveat is that the first term in the LHS of Eq.~\eqref{eq: WTI loop}, $\omega M^0_\t{loop}(k;p;q)$, can be of order unity in
$\omega$ when optical resonance is involved in the loop diagram $M^0_\t{loop}(k;p;q)$. This is because a small change of the frequency $\omega$ can switch retarded and advanced components of the Green's function $G^{R/A}$ in the loop diagram $M^0_\t{loop}(k;p;q)$.
Specifically, the loop diagram can involve terms like $G(i\Omega' + i\Omega) G(i\Omega')$ in the imaginary time formalism, for which we take the analytic continuation with $i\Omega' \to \omega'$ and $i\Omega \to \omega$.
For a small imaginary frequency shift of $\omega=-2i\gamma$ and an optical transition between states $a$ and $b$, we obtain a term like  
\begin{align}
    \frac{1}{(\omega' - \epsilon_{ba}-i\gamma)(\omega' - \epsilon_{ba}+i\gamma)} &\simeq \frac{\pi}{\gamma} \delta(\omega' - \epsilon_{ba})
\end{align}
from  $G(i\Omega' + i\Omega) G(i\Omega')$ after analytic continuation.
Such a term appears due to the branch cut in the Green's function $G(\omega)$ in the real axis of $\omega$.
Namely, when the small frequency shift hits the branch cut of $G(\omega)$, which can happen in the resonant case, the retarded and advanced components are switched in $G(i\Omega' + i\Omega) G(i\Omega')$ after analytic continuation, leading to a term proportional to $1/\gamma$ and a $\gamma$-independent contribution for $\omega M^0_\t{loop}(k;p;q)$.
Indeed, this is the reason why the three-point correlation function describing the usual shift current, which of course involves optical excitation, does not vanish even though it contains an output current vertex with the small frequency $\delta\omega=\omega_1+\omega_2=2i\gamma$ (for DC current response). 

The $O(\delta\omega)$ contribution of the bubble diagram for off-resonant low-frequency responses, which is dictated by the WT identity, is related to electric polarization generation in the system.
Generally, the current $J$ and the electric polarization $P$ are related as $J=dP/dt$, which implies $J(\delta \omega)=-i\delta \omega P(\delta \omega) $ for the Fourier components.
Thus, the current response proportional to $\delta \omega$ means that the electric polarization is induced in the system.
Indeed the diagram $B' \simeq B/(-i\delta\omega)$ in Eq.~\eqref{eq: B'} corresponds to the electric polarization $P_2$ that is carried by a phonon excitation [Eq.~\eqref{eq: P1 P2}]. 
Also, $A(\omega=0)$ means a linear coupling of polarization of the phonon to the external electric field.
Thus, the fact that the phonons carry a shift current with their electric polarization is coherent with the WT identity in the off-resonant situation.

\section{Application \label{sec: app}}
As an application of the present formulation, we consider the phonon shift current in the Rice-Mele model \cite{RiceMele82} coupled with phonons.

We consider the following Hamiltonian
\begin{align}
    H&=\sum_{i} [(1/2)(t+(-1)^i\delta t)(c_{i+1}^\dagger c_{i}+c_{i}^\dagger c_{i+1})
    +(-1)^i m c_{i}^\dagger c_{i}] \n
    &+g_0 \sum_i(-1)^i c_{i}^\dagger c_{i} (a_0 + a_0^\dagger), 
\end{align}
with the hopping amplitude $t$, the hopping alternation $\delta t$, the staggered potential $m$,  and the electron-phonon coupling strength $g_0$.
Here, $c_i$ is the annihilation operator of an electron at site $i$ and $a$ is the annihilation operator of a phonon mode of $q=0$.

In the momentum space representation, we can write the Hamiltonian in a two by two form with Pauli matrices $\bm \sigma_i$ as
\begin{align}
    H(k)&=\sum_{k} \Psi_k^\dagger [t \cos k \sigma_x 
    + \delta t \sin k \sigma_y
    +m \sigma_z] \Psi_k
    \n
    &+\sum_{k} g_0 \Psi_k^\dagger \sigma_z \Psi_k (a_0 + a_0^\dagger),
    \end{align}
where the basis consists of two sites in the unit cell (even and odd $i$) with $\Psi_k =(c_{\t{even},k}, c_{\t{odd},k})^T$.

\begin{figure}[t]
\begin{center}
\includegraphics[width=\linewidth]{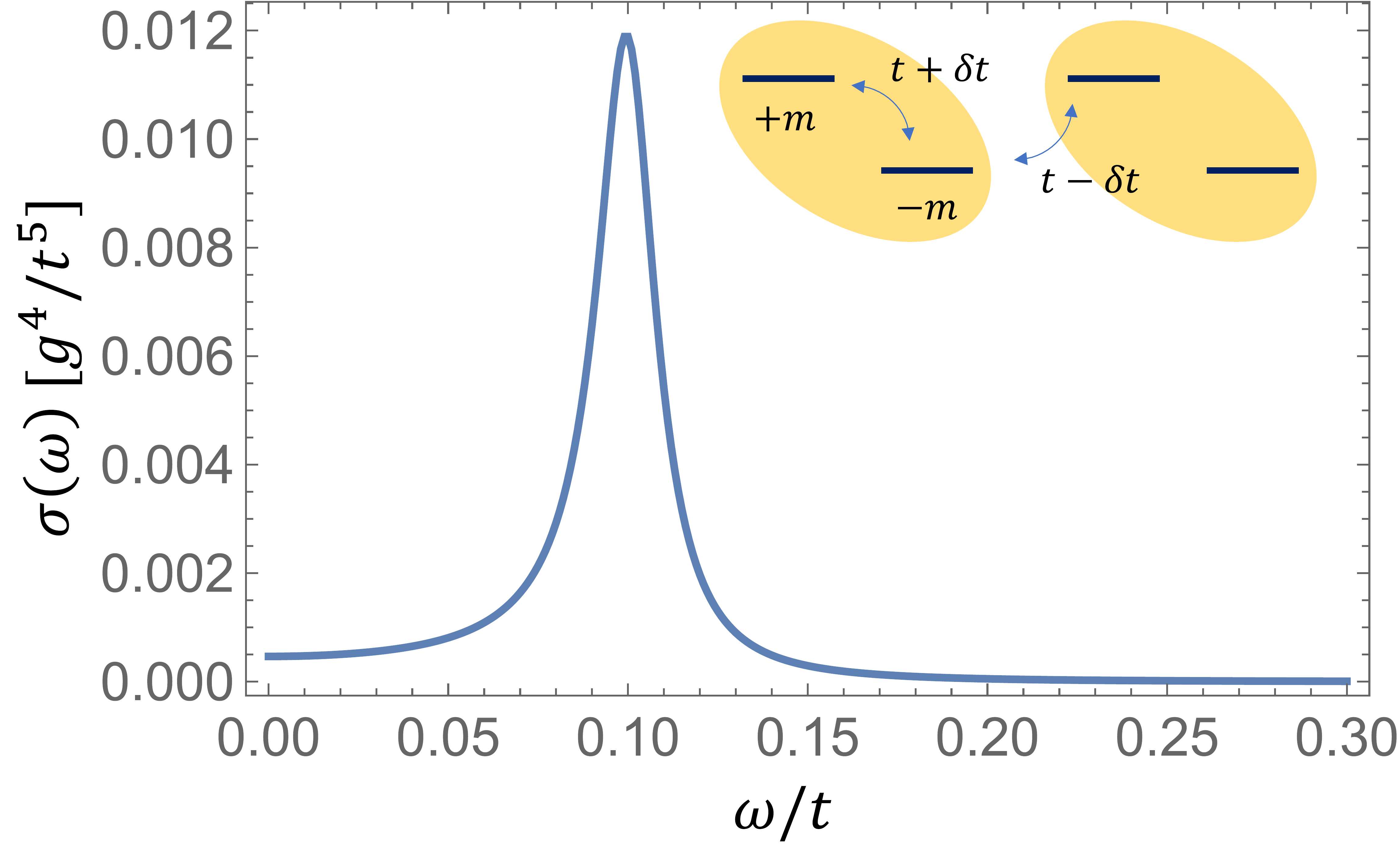}
\caption{\label{fig: conductivity}
Phonon shift current conductivity $\sigma^\t{ph}(\omega)$.
The inset is a schematic for the Rice-Mele model.
We used parameters, 
$\delta t/t=0.3, m/t=-1.5, \omega_0/t=0.1, \gamma/t=0.01$.}
\end{center}
\end{figure}

We show the phonon shift current conductivity $\sigma^\t{ph}(\omega)$ in Fig.~\ref{fig: conductivity}.
$\sigma^\t{ph}(\omega)$ shows a peak structure at the phonon resonance $\omega=\omega_0$,
clearly indicating that the phonon excitation indeed induces photocurrent generation below the electronic band gap.
In the low frequency region below the phonon resonance, the tail of the peak structure gives rise to a nonzero phonon shift current conductivity. 
Since energy dissipation at low frequencies below the phonon resonance generally corresponds to dielectric loss in ferroelectric materials, the nonzero $\sigma^\t{ph}(\omega)$ in this region indicates that the shift current is also induced by the dielectric loss as discussed in Ref.\cite{DielectricLoss24}.

\section{Discussion \label{sec: discussion}}

In this study, we demonstrate that the WT identity indicates that the off-resonant electron loop diagram with a dc current vertex generally gives a contribution proportional to the relaxation rate $\gamma$, which essentially describes the response of electric polarization. This shows that the phonon shift current does not arise from usual charge transport of electron and holes acting as current carriers; instead, it is driven by the electric polarization of the photoexcited phonons themselves. This unified view shows that the phonon shift current can be consistently described within the framework of an effective theory for phonons incorporating their polarization responses.

Let us consider the magnitude of the shift current response by the phonon excitation, compared to that by the usual electronic excitation.
The expression for the phonon shift current $\sigma^\t{ph}(\omega)$ in Eq.~\eqref{eq: sigma ph} is proportional to $g^4$. The phonon excitation takes place at the rate proportional to $g^2$ according to Fermi's golden rule, and each phonon accompanies electric polarization $\propto g^2$ as described by the electron loop $B$ appearing in the self energy for phonons.
Thus overall dependence of phonon shift current on $g$ is $\propto g^4$.
Since the electron-phonon coupling enters in the form of $g/\sqrt{V}$ in the Hamiltonian, $g/\sqrt{V_c}$ ($V_c$: unit cell volume) has the dimension of energy.
If we write the typical energy scale of the electronic band gap as $E_g$, the typical strength of $g$ is characterized by $g^2/(V_c E_g) \sim \omega_0$ \cite{Okamura22, Sun22}.
Since the delta function $\delta(\omega-\omega_0)$ gives a factor of the density of state (inversely proportional to corresponding energy scales), the phonon shift current proportional to $g^4$ indicates that the ratio between the phonon shift current $\sigma^\t{ph}$ and the electronic shift current $\sigma^\t{el}$ is roughly given by \begin{align}
    \frac{\sigma^\t{ph}}{\sigma^\t{el}} \simeq \frac{g^4}{V_c^2 E_g^4}\left(\frac{\omega_0}{E_g}\right)^{-1} \sim \frac{\omega_0}{E_g}.
    \label{eq: sigma ratio}
\end{align} 
The electronic bandwidth (band gap) is of the order of 1 eV, while the phonon energy ranges from a few meV for soft phonons to $\sim 100$ meV for optical phonons.
Thus, the phonon shift current is expected to be $1 \sim 2$ orders of magnitude smaller than the electronic shift current.
The phonon shift current can be comparable to the electronic one for semimetals or narrow band gap semiconductors.

The diagram in Fig.~\ref{fig: diagrams} gives a contribution to photocurrent at the phonon resonance that is independent of the relaxation rate $\gamma$. Here, the product of phonon propagators $D(\omega+i\gamma)D(\omega-i\gamma)$ (each corresponding to retarded and advanced Green's function for the phonons) gives the factor with a delta function, $(\pi/\gamma)\delta(\omega-\omega_0)$, at the phonon resonance. Combining this with the the $\gamma$ linear term in
$B(\omega)$ gives a $\gamma$ independent contribution as in Eq.~\eqref{eq: phonon shift current}.
Generally, under the constraint of the WT identity, the nonresonant bubble diagram that includes the dc current output gives a contribution proportional to $\gamma$ as shown in Sec.~\ref{sec: WT id}. In order to obtain an overall contribution of order unity with respect to $\gamma$, it is necessary to attach two resonant phonon propagators ($D(\omega+i\gamma)$ and $D(\omega-i\gamma)$) on both sides of the nonresonant bubble diagram containing the dc current output.
Since connecting these two resonant phonon propagators to each electron bubble brings an overall factor of $g^4$, we can see that only the diagram in Fig.~\ref{fig: diagrams}(a) yields the $O(\gamma^0)$ contribution at order $g^4$ that shows a resonance at the phonon energy.
While there exist other contributions with the electron-phonon coupling to $\sigma^\t{ph}(\omega)$ such as those containing Debye–Waller terms modifying the energy of the virtual electron–hole pair excitation\cite{Antonius22},
those contributions become higher order terms in either $g$ or $\gamma$.

Also, the diagrams previously studied in Ref.~\cite{DielectricLoss24} for phonon shift current give a subleading contribution proportional to $\gamma$. 
There, two electron bubble diagrams connected by one phonon propagator were considered. One electron bubble diagram is given by $A(\omega)$, whereas the other electron bubble diagram (which we call $\tilde B$) is obtained by replacing electron-phonon vertex with a current vertex of incoming light frequency $\omega_2$ in $B(\omega)$ (Fig.~2(d) in Ref.~\cite{DielectricLoss24}).
As detailed in Sec.~\ref{sec: WT id}, the WT identity indicates an electron bubble diagram with a current vertex with zero frequency transfer $\omega_1+\omega_2=0$ becomes zero when the electron line is off-resonant. 
Since the frequency $\omega$ of the external electric field is off-resonant with the electronic excitation, such an electron loop diagram $\tilde B$ also undergoes a cancellation and its contribution becomes $O(\gamma)$. Thus the previously considered contribution in \cite{DielectricLoss24} becomes proportional to $\gamma$ and is subleading (which we write $\sigma^\t{sub}$), 
whereas the leading contribution independent of $\gamma$ is given by the present diagram in Fig.~\ref{fig: diagrams}. 
We note that the current vertex $\partial_k g$ originating from phonon modulated hopping was not incorporated in Ref. \cite{DielectricLoss24} and the cancellation due to the WT identity was not fully recognized there.
Meanwhile, the present contribution (Fig.~\ref{fig: diagrams}) still gives rise to the photocurrent generation by phonon excitations and that by the dielectric loss in the low frequency region, as originally discussed in Ref.~\cite{DielectricLoss24}. 
Also the subleading contribution $\sigma^\t{sub}$ is proportional to $g^2 \gamma$ and is estimated in comparison with the electronic shift current $\sigma^\t{el}$ as $\sigma^\t{sub}/\sigma^\t{el}\simeq (g^2/V_c E_g^2)(\gamma/E_g) (\omega_0/E_g)^{-1} \simeq \gamma/E_g$.
Comparing this to Eq.~\eqref{eq: sigma ratio}, the ratio of the two contributions to the phonon shift current is roughly given by $\sigma^\t{sub}/\sigma^\t{ph} \simeq \gamma/\omega_0$.
Thus $\sigma^\t{ph}$ is dominant over $\sigma^\t{sub}$ for $\omega_0 \gg \gamma$, whereas $\sigma^\t{ph}$ and $\sigma^\t{sub}$ can become comparable for $\omega_0 \sim \gamma$.
Typically the relaxation rate is $\gamma \sim$10 meV and the optical phonon frequency is $\omega_0 \sim$100 meV, where the present term $\sigma^\t{ph}$ becomes dominant over $\sigma^\t{sub}$. In contrast, for the case of soft phonons that appear near the structural phase transition, the phonon frequency reduces to $\omega_0 \sim$10 meV, where $\sigma^\t{ph}$ and $\sigma^\t{sub}$ can become comparable.

\begin{acknowledgments}
We thank Youtarou Takahashi, Yoshihiro Okamura, and Jiaming Hu for fruitful discussions.
T.M. was supported by 
JSPS KAKENHI Grant 23K25816, 23K17665, and 24H02231.
N.N. was supported by JSPS KAKENHI Grant Numbers 24H00197, 24H02231 and 24K00583.
N.N. was supported by the RIKEN TRIP initiative.
\end{acknowledgments}

\appendix

\section{Expressions for diagrams $A(\omega)$ and $B(\omega)$ \label{app: B(w)}}
In this section, we present the derivation of expressions for $A(\omega)$ and $B(\omega)$ in  Eq.~\eqref{eq: A1} and Eq.~\eqref{eq: B(w)}.

In the imaginary time formulation, the contributions of the diagrams in Fig.~\ref{fig: diagrams}(b) are written as
\begin{align}
    &A(i\Omega_1) = \n
    &\frac{1}{i(i\Omega_1)} \int\frac{d\Omega}{2\pi} \int[dk] \t{Tr}\Bigg[v^\beta G(k, i\Omega + i \Omega_1) g G(k, i \Omega) \n
    &+\partial_{k_\beta} g G(k,i\Omega) \Bigg].
\end{align}
Integrating the second term by parts, we obtain
\begin{align}
    A(i\Omega_1) &= \frac{1}{i(i\Omega_1)} \int\frac{d\Omega}{2\pi} \int[dk] \t{Tr}\Bigg[v^\beta G(k, i\Omega + i \Omega_1) g G(k, i \Omega) \n
    &-v^\beta G(k, i\Omega) g G(k, i \Omega) \Bigg].
\end{align}
Using 
\begin{align}
    I_{2,ab}(i\Omega_1)&=
    \int \frac{d\omega}{2\pi} \frac{1}{(i\omega+i\Omega_1-\epsilon_b)(i\omega-\epsilon_a)}\n
    &=\frac{f_{ab}}{i\Omega_1-\epsilon_{ba}},
\end{align}
and performing the analytic continuation $i\Omega_1 \to \omega+i\gamma$,
we can express $A(\omega)$ with matrix elements as
\begin{align}
    A(\omega) &= \frac{1}{i\omega} \int [dk]v^\beta_{ab}g_{ba}[I_{2,ab}(\omega+i\gamma)-I_{2,ab}(0)]\n
    &=\int [dk]\frac{f_{ab}v^\beta_{ab}g_{ba}}{(i\epsilon_{ba})(\omega-\epsilon_{ba})},
\end{align}
which leads to Eq.~\ref{eq: A1}.
Note that we neglected $\gamma$ by assuming the off-resonant condition, $|\omega-\epsilon_{ba}| \gg \gamma$.

The contributions of the diagrams in Fig.~\ref{fig: diagrams}(c) are written with the imaginary time Green's functions as
\begin{align}
&B(i\Omega_1, i\Omega_2) = \n
&\int \frac{d\Omega}{2\pi} \int[dk] \t{Tr}\Bigg[\partial_{k_\alpha} g G(k, i\Omega + i \Omega_2) g G(k, i \Omega) \n
&+\partial_{k_\alpha} g G(k, i\Omega + i \Omega_1) g G(k, i \Omega) \n
&+v^\alpha G(k, i\Omega + i\Omega_1 + i \Omega_2) g G(k, i\Omega + i \Omega_2) g G(k, i \Omega) \n
&+v^\alpha G(k, i\Omega + i\Omega_1 + i \Omega_2) g G(k, i\Omega + i \Omega_1) g G(k, i \Omega)
\Bigg],
\end{align}
where the first two terms are bubble diagram contributions and the last two are triangle diagram contributions.
With integrating by parts, the first term can be rewritten as 
\begin{align}
    &\int[dk] \t{Tr}[\partial_{k_\alpha} g G(k, i\Omega + i \Omega_2) g G(k, i \Omega)] \n
    &=
    -\int[dk] \t{Tr}\Bigg[g G(k, i\Omega + i \Omega_2) v^\alpha G(k, i\Omega + i \Omega_2) g G(k, i \Omega) \n
    &+g G(k, i\Omega + i \Omega_2) \partial_{k_\alpha} g G(k, i \Omega) \n
    &+g G(k, i\Omega + i \Omega_2) g G(k, i \Omega) v^\alpha G(k, i\Omega) \Bigg],
\end{align}
Here we used the identity $\partial_{k_\alpha} G(\omega)=G(\omega) v^\alpha G(\omega)$.
By shifting the integral variable $\Omega$ and reordering terms in the trace,
we obtain
\begin{widetext}
\begin{align}
B(i\Omega_1, i\Omega_2)
&=\int \frac{d\Omega}{2\pi} \int[dk] \t{Tr}\Bigg\{ \partial_{k_\alpha} g [G(k, i\Omega + i \Omega_1)-G(k, i\Omega -  i \Omega_2)] g G(k, i \Omega) \n
&+v^\alpha [G(k, i\Omega + i \Omega_1+i \Omega_2)-G(k, i\Omega ))] g G(k, i \Omega+i \Omega_2) g G(k, i\Omega) \n
&+  v^\alpha [G(k, i\Omega+i\Omega_1+i\Omega_2) g G(k, i \Omega+i\Omega_1) - G(k,i\Omega)g G(k,i\Omega-i\Omega_2)] g G(k,i\Omega)\Bigg\} .
\end{align}
We perform the $\Omega$ integral and the analytic continuation, $i\Omega_1 \to \omega_1$, $i\Omega_2 \to \omega_2$,
which leads to 
\begin{align}
B(\omega_1, \omega_2) &=
\int[dk] \Bigg\{
\sum_{a,b}(\partial_{k_\alpha}g)_{ab} g_{ba} 
[I_{2,ab}(\omega_1)-I_{2,ab}(-\omega_2)] \n
&+
\sum_{a,b,c} v^\alpha_{ac} g_{cb} g_{ba} [I_{3,abc}(\omega_2,\omega_1)-I_{3,abc}(\omega_2,-\omega_2)] \n
&+ \sum_{a,b,c} v^\alpha_{ac} g_{cb} g_{ba}  [I_{3,abc}(\omega_1,\omega_2)-I_{3,abc}(-\omega_2,\omega_2)]
\Bigg\},
\end{align}
with
\begin{align}
I_{2,ab}(i\Omega_1)
&=\int \frac{d\Omega}{2\pi} \frac{1}{(i\Omega+i\Omega_1-\epsilon_b)(i\Omega-\epsilon_a)} \n
&=\frac{f_{ab}}{i\Omega_1-\epsilon_{ba}} ,
\end{align}
and
\begin{align}
& I_{3,abc}(i\Omega_1,i\Omega_2) \n
&=\int \frac{d\Omega}{2\pi} \frac{1}{(i\Omega+i\Omega_1+i\Omega_2-\epsilon_c)(i\Omega+i\Omega_1-\epsilon_b)(i\Omega-\epsilon_a)} \n
&=\frac{1}{\epsilon_{ac}+i\Omega_1+i\Omega_2}\left(\frac{f_{ab}}{i\Omega_1-\epsilon_{ba}} - \frac{f_{cb}}{-i\Omega_2 -\epsilon_{bc}}\right).
\end{align}
Substituting $\omega_1=\omega+i\gamma$, $\omega_2=-\omega+i\gamma$,
we obtain 
\begin{align}
B(\omega+i\gamma, -\omega+i\gamma) &= \gamma B'(\omega) +O(\gamma^2),
\end{align}
with
\begin{align}
B'(\omega) &= 
2i\int[dk] \Bigg\{
\sum_{a,b}(\partial_{k_\alpha}g)_{ab} g_{ba} 
\partial_\omega I_{2,ab}(\omega) \n
&+
\sum_{a,b,c}v^\alpha_{ac} g_{cb} g_{ba} [\partial_{\omega'}I_{3,abc}(-\omega,\omega')|_{\omega'=\omega} 
+ \partial_{\omega'}I_{3,abc}(\omega',-\omega)|_{\omega'=\omega}]
\Bigg\},
\label{eq: B' app}
\end{align}
which indicates that the leading order contribution is proportional to $\gamma$.
\end{widetext}

\bibliography{ref}% Produces the bibliography via BibTeX.

\end{document}